\documentclass[conference]{IEEEtran}
\IEEEoverridecommandlockouts
\usepackage{cite}
\usepackage{amsmath,amssymb,amsfonts}

\usepackage{graphicx}
\usepackage{textcomp}
\usepackage{xcolor}
\AtBeginDocument{%
  \providecommand\BibTeX{{%
    Bib\TeX}}}
\usepackage{comment}

\usepackage{amsmath}
\usepackage{eqlist} 
\usepackage[nosepfour,warning,np,debug,autolanguage]{numprint}
\usepackage{acro}
\usepackage{booktabs}
\usepackage{graphicx}
\usepackage[most]{tcolorbox}
\usepackage{titlesec}
\usepackage{hyperref}
\usepackage{multirow}
\usepackage{siunitx}
\usepackage{subcaption}
\usepackage{graphicx}
\usepackage{svg}
\usepackage{booktabs}
\usepackage{xcolor}
\usepackage{textcomp}

\usepackage{algorithm}
\usepackage[noend]{algpseudocode}
\usepackage{float} 
\usepackage{array}
\usepackage{soul}
\usepackage{tikz}
\usepackage{mathtools}
\usepackage{cleveref}
\newcounter{mysubequations}

 \makeatletter
    \newcommand{\linebreakand}{%
      \end{@IEEEauthorhalign}
      \hfill\mbox{}\par
      \mbox{}\hfill\begin{@IEEEauthorhalign}
    }
    \makeatother

\def\BibTeX{{\rm B\kern-.05em{\sc i\kern-.025em b}\kern-.08em
    T\kern-.1667em\lower.7ex\hbox{E}\kern-.125emX}}
\begin{document}

\title{GraphEx: A Graph-based Extraction Method for Advertiser Keyphrase Recommendation}

\author{\IEEEauthorblockN{Ashirbad Mishra}
\IEEEauthorblockA{\textit{Dept. of Computer Science} \\
\textit{Pennsylvania State University}\\
University Park, PA, USA \\
amishra@psu.edu}
\and
\IEEEauthorblockN{Soumik Dey}
\IEEEauthorblockA{\textit{eBay Advertising} \\
\textit{eBay Inc.}\\
San Jose, CA, USA \\
sodey@ebay.com}
\and
\IEEEauthorblockN{Hansi Wu}
\IEEEauthorblockA{\textit{eBay Advertising} \\
\textit{eBay Inc.}\\
San Jose, CA, USA \\
hanswu@ebay.com}
\and
\IEEEauthorblockN{Jinyu Zhao}
\IEEEauthorblockA{\textit{eBay Advertising} \\
\textit{eBay Inc.}\\
San Jose, CA, USA \\
jinyzhao@ebay.com}
\linebreakand
\IEEEauthorblockN{He Yu}
\IEEEauthorblockA{\textit{eBay Advertising} \\
\textit{eBay Inc.}\\
Shanghai, China \\
hyu1@ebay.com}
\and
\IEEEauthorblockN{Kaichen Ni}
\IEEEauthorblockA{\textit{eBay Advertising} \\
\textit{eBay Inc.}\\
Shanghai China \\
kani@ebay.com}
\and
\IEEEauthorblockN{Binbin Li}
\IEEEauthorblockA{\textit{eBay Advertising} \\
\textit{eBay Inc.}\\
San Jose, CA, USA \\
binbli@ebay.com}
\and
\IEEEauthorblockN{Kamesh Madduri}
\IEEEauthorblockA{\textit{Dept. of Computer Science} \\
\textit{Pennsylvania State University}\\
University Park, PA, USA \\
madduri@psu.edu}
}

\maketitle

\begin{abstract}
Online sellers and advertisers are recommended keyphrases for their listed products, which they bid on to enhance their sales. One popular paradigm that generates such recommendations is Extreme Multi-Label Classification (XMC), which involves tagging/mapping keyphrases to items. We outline the limitations of training XMC models on click data for keyphrase recommendations on E-Commerce platforms. We introduce \textit{GraphEx}, an innovative graph-based approach that recommends keyphrases to sellers using extraction of token permutations from item titles. Additionally, we demonstrate traditional metrics such as precision/recall isn't reliable on click-based data in practical applications, thereby necessitating a robust framework to evaluate performance in real-world scenarios. Our evaluation is designed to assess the relevance of keyphrases to items and the potential for buyer outreach. GraphEx outperforms production models at eBay, achieving the objectives mentioned above. It supports near real-time inferencing in resource-constrained production environments and scales effectively for billions of items.
\end{abstract} 


\begin{IEEEkeywords}Keyphrase Recommendation, Sponsored search advertising, Graph Algorithms, Efficient Scalable Processing.
\end{IEEEkeywords}


\maketitle

\section{Introduction}
\label{s:chap4_introduction}
\begin{figure*}[h]
\centering

\begin{subfigure}{8.5cm}
\centering
\includegraphics[width=\linewidth]{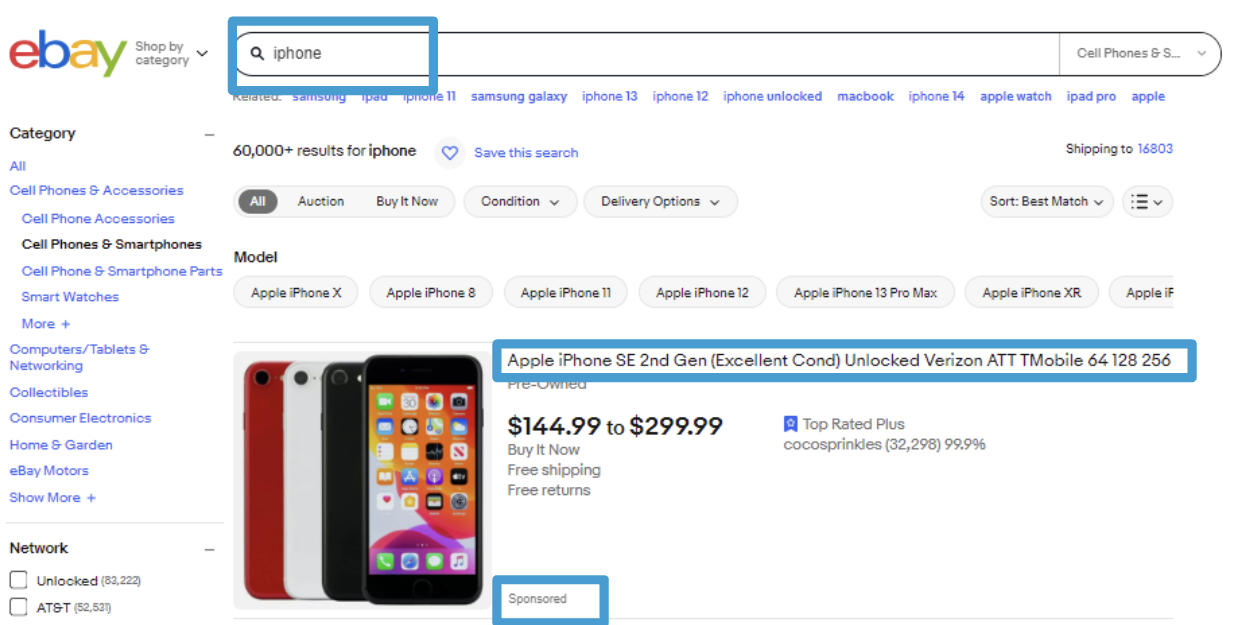}
  \caption{Buyer side}
  \label{fig:1a}
\end{subfigure}\qquad
\begin{subfigure}{8.5cm}
\centering

\includegraphics[width=\linewidth]{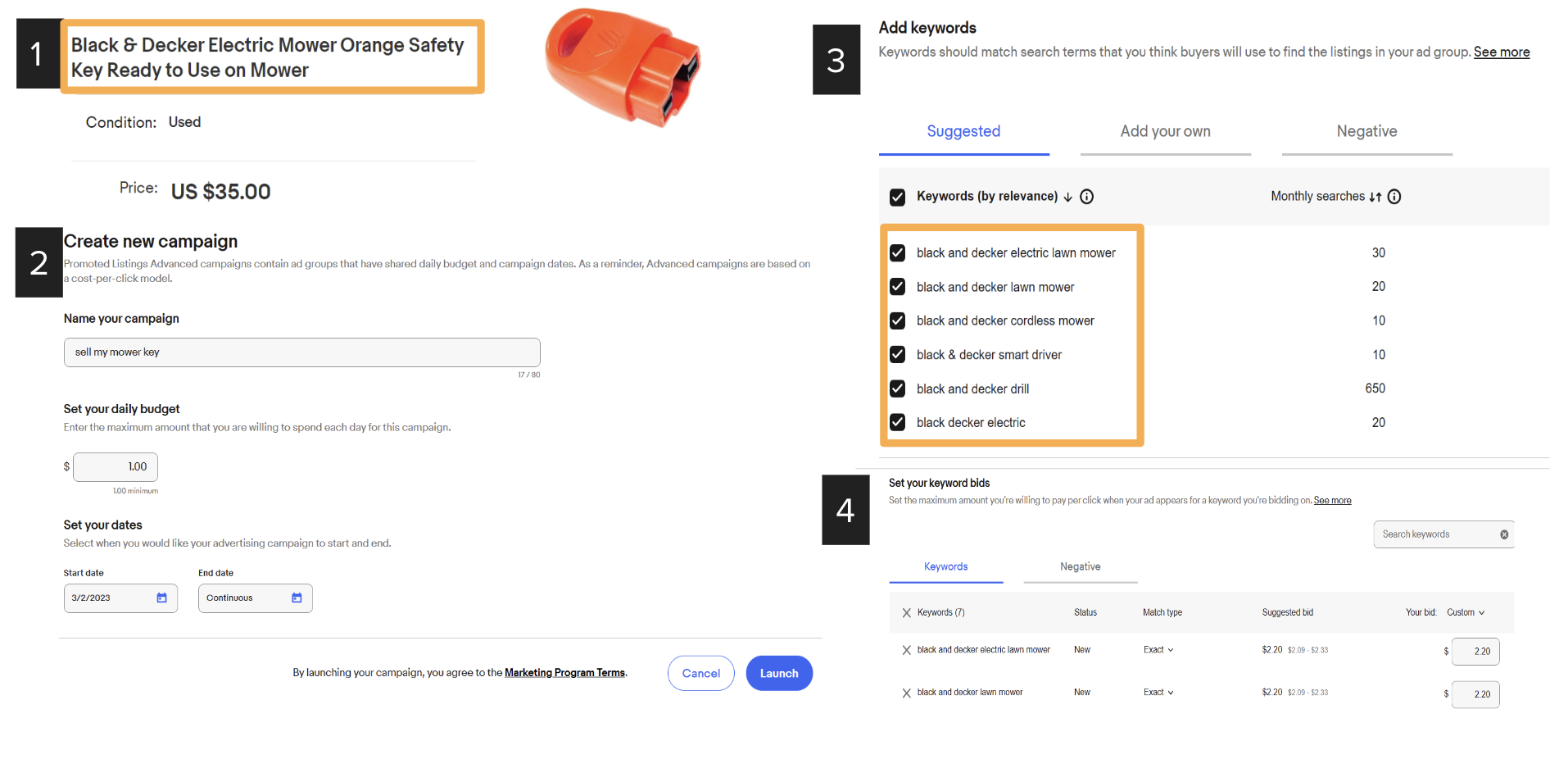}
\caption{Seller Side}
\label{fig:1b}
\end{subfigure}
\caption{Screenshot of our keyphrases for manual targeting in Promoted Listings Priority\textsuperscript{TM} for eBay Advertising.}
\label{fig:screenshot}
\vspace{-4mm}
\end{figure*}

In the online e-commerce advertisement space,
\textit{keyphrase recommendations} are offered to sellers/advertisers who want to bid on buyers/users' search queries for a better placement of their inventory on the search result pages (SRP) which increases the item's engagement. Keyphrases\footnote{We term buyer search queries as \textit{keyphrases} and use queries and \textit{keyphrases} interchangeably.} are generally recommended as shown in Figure~\ref{fig:1b} on the right in real time for the items if they are relevant to them. The keyphrase recommendations provided by Advertising are then matched to actual queries by eBay Search and enter auctions where the sellers/advertisers with the highest bid wins the auction and gain a \textit{prominent sponsored placement} on the SRP page. Hence, it is important to suggest keyphrases that are an exact match to the queries to prevent missed targeting. When buyers interact with the SRP by searching for a query and clicking on an item as shown in Figure~\ref{fig:1a}, the advertiser or seller is charged (CPC or cost-per-click business model) for their advertised item. Any interaction (click, add to cart, buy etc.) on the SRP page is logged in the search logs. An association between the search queries or keyphrases and items can be derived which we call as \textit{click-based ground truths}.

This problem of keyphrase recommendation has been formulated as an Extreme Multi-Label Classification (XMC) problem and well studied in  ~\cite{dahiya2023ngame,renee_2023,Dahiya23b,you2019attentionxml,dahiya2021deepxml, ashirbad-etal-2024}. The data for training the XMC tagging models is generally the click-based ground truths sourced from search logs. There are multiple challenges impacting this research area which we describe in detail.


\subsection{Challenges}
\label{ss:challenges}

\subsubsection{Budgeted Recommendation}
\label{sss:bud_reco}
\vspace{1mm}
Modern recommendation systems operate on a budget of recommendations, i.e. a maximum number of recommendations (which in eBay's case is 1000 keyphrases per adgroup with an adgroup containing at max 1000 items). Within this constrained budget@k \cite{PUSL} the targeting significance of every keyphrase becomes paramount. Keyphrases can be classified as head or tail keyphrases according to their search frequency. Head keyphrases are generally less in number but searched frequently by buyers. Targeting such head keyphrases leads to increased revenue since more buyers are inclined to search for them, resulting in more clicks and more buys. XMC models are agnostic towards head or tail keyphrases and end up focusing on recommending the more copious tail keyphrases \cite{Dahiya23b, dahiya2023ngame} while on a budget --- missing out on potentially more important head keyphrases.

\vspace{2mm}
\subsubsection{Click data biases}
\label{sss:click_biases}
\vspace{1mm}
The large set of labels or keyphrases in the click data challenges exhaustive annotation,  resulting in missing labels due to sparsity of dataset (i.e. 96\% of items don't have clicks associated with them) and the clicks of items are influenced by various biases like popularity bias, exposure bias~\cite{surveybias}, sample-selection bias~\cite{sampleselectionbias} etc. While clicks can be treated as reliable labels for relevance, the absence of clicks cannot be taken as a sign of irrelevance~\cite{MNAR}.

While these biases have been discussed in ~\cite{clicksNquery,beyondPosBias,learning2rank,debiasedness}, we contextualize them in this domain of advertiser keyphrase recommendation. When buyers are shown items based on a specific query or keyphrase, the way the presented items are ranked can introduce bias, influencing buyer engagement. This biased ranking suggests that an item lacking clicks or sales for a given keyphrase isn't necessarily irrelevant to that keyphrase~\cite{lim2015top}. Instead, it might be less popular and thus lower ranked, leading buyers to overlook it (exposure and popularity biases). These unpopular items need the help of advertising to level the playing field by promoting their items to a favorable rank, increasing their visibility and engagement. In explicit feedback data, such as click-based data, signals can be Missing-not-at-Random (MNAR) \cite{MNAR}, which means that systems trained solely on these feedback signals are likely to perpetuate these biases in their predictions.

\begin{figure}[h]
    \centering
    \includegraphics[width=0.8\linewidth]{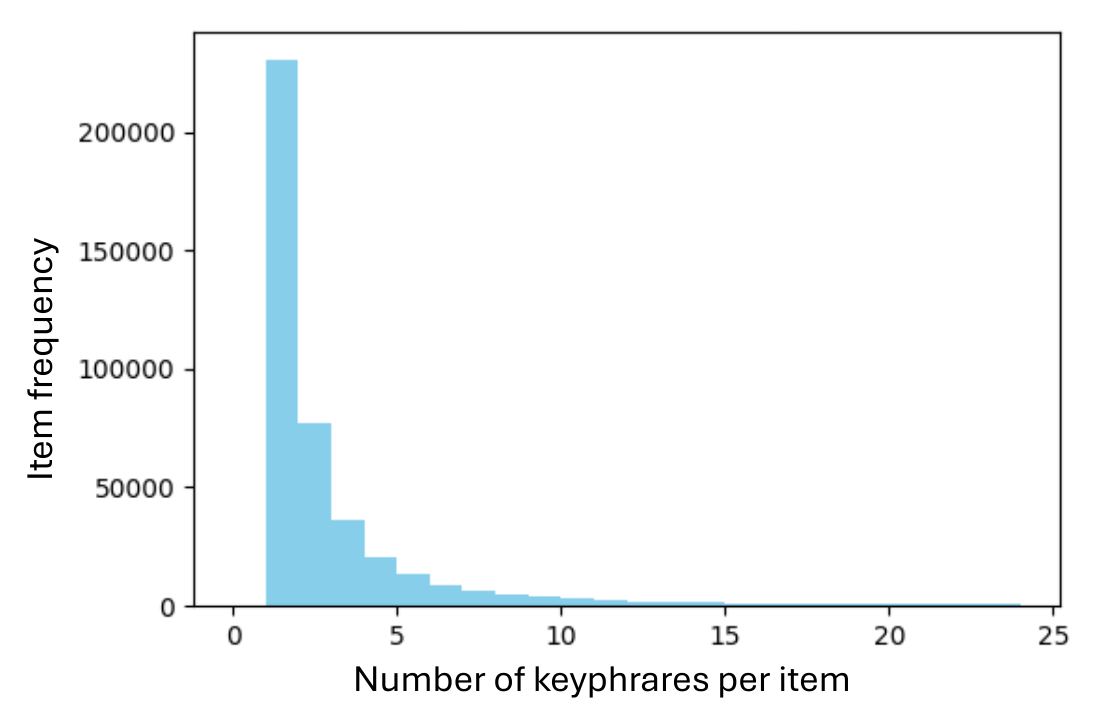} 
    \caption{Distribution of click-data in terms of items and the number of queries associated with them.}
    \label{fig:distribution}
\end{figure}

Even for the items that get a click, 90\% of such items are associated with only one query in terms of clicks ---  as shown in Figure \ref{fig:distribution}, while sellers expect a healthy set of recommendations from us (around 20-30 per item). 

\subsubsection{Click-based Evaluation}
\label{sss:click_eval}
\vspace{1mm}
Sourcing the same click data as the training set for different XMC tagging models perpetuates not only the described biases but also the lack of diversity in the model's recommendations. The offline evaluation of such models is typically done using metrics like \textit{Precision}, \textit{Recall}, \textit{F1} and so on. These metrics facilitate comparison by emphasizing retrieval capability, i.e how well are the models able to retrieve the existing associations of keyphrases with the items.

The diversity issue is exacerbated by evaluating with these metrics  due to two main reasons: \textit{Lack of Ground Truths} and \textit{Model Convergence}. Due to the curation process, there is a lack of ground truths, since most items are excluded from the training set, as evident from Section~\ref{sss:click_biases} and Figure~\ref{fig:distribution}. So, the metrics aren't able to evaluate a model's recommendation beyond the labels gathered per item which are very sparse. This is problematic especially considering the budgetary constraints described in Section~\ref{sss:bud_reco}.

To understand the convergence of the model, let us take an example instance $T$, associated with a ground truth label $k1$ in the training set. All candidate models make certain choices to increase the probability of predicting $k1$ for inputs similar to $T$. This aligns the tagging-based models to predict a similar subset of labels for $T$, thus reducing diversity between the predictions of different models. Even with a $10\%$ increase in the precision/recall scores of subsequent XMC models, the recommendations do not have sufficient diversity to obtain substantial clicks. The impact of models is further dampened by the 100\% recall models in production (database lookup of queries that generated clicks in relation to the items). Models with higher recall will have less impact as they will be de-duplicated against the 100\% recall model's recommendations. 

\vspace{2mm}
\subsubsection{Keyphrase Targeting}
\label{sss:key_target}
\vspace{1mm}
The XMC tagging models are required to be regularly updated (preferably daily) to keep up with the churn of new queries (2\% churn every day) and other factors such as holidays and seasonality. Out of Vocabulary (OOV) models can recommend keyphrases that are absent in the training set --- thus circumventing the need for daily training. 

The OOV models however also suffer from biases in click data as described in~\cite{PUSL}. Data augmentation techniques such as rejection sampling~\cite{PUSL} have been shown to mitigate these biases. As described earlier, the exact matching of recommended keyphrases to the search queries in auctions along with the budgetary constrains from section~\ref{sss:bud_reco} makes OOV model's inaccurate targeting less desirable.

Disengaging from the click-based associations can help mitigate their biases, however the targeting and the evaluation of such models become a challenge.

\vspace{2mm}
\subsubsection{Execution Performance}
\label{sss:exec_perf_chal}
\vspace{1mm}
For e-commerce platforms, a vital necessity is that the recommendations are in real-time or near real-time, so the models should also have inference latency of a few milliseconds. Alternatively, models should have a sufficiently low inference latency for daily batch prediction. Due to sharing of resources and the cost of acquisition, GPUs (50x higher cost than CPUs) and high-memory systems are generally not available. 

This complicates the deployment of models involving LLMs that require high GPU costs while still maintaining a healthy margin for sellers and the platform. Moreover, LLMs have large inference and training times~\cite{kaddour2023} and have problems scaling on large datasets~\cite{MLSYS2023_LLMscaling,xin-etal-2020-deebert} making them unsuitable for deployment in latency-sensitive applications. 

Due to the model refresh requirements from Section~\ref{sss:key_target}, models with smaller training and setup times are absolutely necessary, and minimal tuning is crucial to decrease the engineering effort.

\subsection{Scope and Contributions}
\label{ss:chap4_scope_cont}
In this work, we focus on the challenges mentioned in the previous section. We limit ourselves to retrieving keyphrases based on item's title and belonging to the items' categorical populace under budgetary constraints, especially those keyphrases that are actively and frequently searched by buyers. The extraction is done in an unsupervised setting where the keyphrases for each item are unknown during training. 
In fact, we restrict the curation of keyphrases (more details in~\ref{ss:chap4_datasets}) to include only those that have a high \textit{ search volume} (number of searches made by buyers) based on buyer searches. 
XMC models suffer from the biases mentioned in Section~\ref{sss:click_biases} and as a result tend to recommend tail keyphrases (Section~\ref{sss:bud_reco}). Our distinction is that by using categorical population dynamics of keyphrases, we decouple the keyphrases from item click-based engagement. This allows us to retain the essential bias towards head keyphrases (attractive to advertisers) while getting rid of the negative bias (against non-popular items which are the main target of advertisement) which helps us solve Challenge~\ref{sss:click_biases}.


Our contribution to this work is summarized as follows: 
\begin{itemize}
    \item An innovative graph-based extraction algorithm for keyphrase recommendation that is transparent and easy to interpret. 
    \item The design of the algorithm and the process of data collection have been specifically geared toward mitigating click-based biases while maintaining the advertiser friendly head keyphrase bias under budgetary constraints. 
    \item Provide a new robust framework for the evaluation of incremental impact of recommendation models in terms of relevance and diversity metrics.
    \item A low latency and sustainable model that runs without GPUs and scales for daily inference and training on billions of items.
\end{itemize}

\section{Related Work}
\label{s:chap4_related_work}

Bipartite graphs are extensively utilized across various fields to model user search behaviors from logs, such as query-url \cite{aglo2000QueryLog,baeza2007extracting} and query-ad graphs \cite{anastasakos2009collaborative}. Typically, techniques applied to these bipartite graphs calculate query similarities according to the items they are linked with. This similarity information is subsequently used to propose queries for new items. Simrank++ \cite{antonellis2008simrank++} enhances query similarity measures by reducing the iteration count required for convergence and by adjusting the similarity score with a multiplier that reflects the number of shared neighbors between queries. Nevertheless, in the worst-case scenario, these methods demand a pairwise comparison of all queries (i.e., quadratic complexity), which becomes impractical with a large volume of keyphrases. Additionally, ranking recommended queries linked to similar items based on their relevance presents another challenge.

SOTA models for extreme multi-label classification (XMC)~\cite{dahiya2023ngame,renee_2023,Dahiya23b,you2019attentionxml,dahiya2021deepxml} predominantly leverage deep neural networks (DNNs) and commonly employ one-vs-all (OVA) classifiers. Although certain models~\cite{dahiya2023ngame,Dahiya23b} need label features, others such as \emph{AttentionXML}~\cite{you2019attentionxml}, \emph{Renee}~\cite{renee_2023}, and \emph{DeepXML/Astec}~\cite{dahiya2021deepxml} do not. Among the DNN models assessed in~\cite{Bhatia16}, DeepXML/Astec~\cite{dahiya2021deepxml} demonstrates scalability to large datasets and achieves a relatively short training duration compared to rival methods. However, \cite{ashirbad-etal-2024} shows the non-viability of Astec and AttentionXML on large categories on eBay and the cost and scalability challenges of GPU inference associated with it.

Other efficient XMC tagging models include --- \textit{fastText}~\cite{fastText1,fastText2} which is an effective CPU-based option for managing extensive workloads. fastText creates word embeddings using the \emph{CBOW} model and employs a straightforward linear neural network model with hierarchical softmax to improve the efficiency of training and inference processes. A key reason for fastText's effectiveness is its integration of \emph{subword} information into the embeddings. The size of the model can be easily reduced to conserve storage using methods such as quantization~\cite{fasttext3}, as well as pruning the vocabularies of important phrases and title words. \textit{Graphite}~\cite{ashirbad-etal-2024} is another SOTA XMC model that uses bipartite graphs to map words/tokens to the data points and then map them to the labels associated with the data points. It is implemented for multi-core systems having infinitesimal training time and uses parallelization for real-time inferencing. It's training and inference scales well for hundreds of millions of items and labels. \textit{SL-emb}~\cite{Wang2021PersonalizedEE} uses embeddings of the item's title to compare and find similar listings, and then recommend the related queries. SL-emb is a
dense retrieval model whose inference is implemented in two stages, namely, embedding generation and ANN\cite{hnsw}. The SL-emb model does not need to be trained daily and is based on the hypothesis that semantically close items have similar keyphrases. The SL-emb model is trained on Recs data (similar item recommendation for a hero item), which is shown to mitigate some of the bias from Ads click data \cite{rec4ad}.

Rule-based heuristic models are also in production as simple models that can provide recommendations for existing popular items. \textit{Rules Engine (RE)} is a simple technique that stores item-keyphrase associations based on their co-occurrences (associated with buyer activity) in the search logs during the last 30 days which is around 13\% of all active items (item coverage). It recommends keyphrases only for items in which buyers have shown interest and not for any new items. This is a 100\% recall model in which buyers' interest is reflected back to them. \textit{SL-query} is also a rule-based model based on the hypothesis that similar listings share similar queries. SL-query recommends the associated queries of listings that share a keyphrase with the seed item. Both SL models' predictions are truncated from a higher number of predictions using a Jaccard coefficient~\cite{jaccard} threshold to ensure relevance of the predictions. The RE and SL-query models have a low item-coverage (since items with query associations are quite sparse as described in Section \ref{sss:click_biases}) and don't offer recommendations on cold items.  The implementation of the RE and SL techniques also employs a few other details, which we cannot discuss due to proprietary constraints. 

Keyphrase generation via open-vocabulary models like GROOV \cite{simig-etal-2022-open}, One2Seq \cite{one2seq,one2set} and One2One \cite{one2one1,chen-etal-2019-integrated,Chen_Gao_Zhang_King_Lyu_2019} are susceptible to recommending keyphrases that are not part of the label space \cite{PUSL}. Another formulation for keyphrase recommendation is keyphrase extraction with methods such as \textit{keyBERT}\cite{grootendorst2020keybert}, which have conventionally treated keyphrase recommendation as a two-step problem: keyphrase generation and keyphrase ranking. The basic keyBERT module considers keyphrase generation as an n-gram-based permutation problem, i.e., it generates all possible n-grams for a given n-gram range. The keyphrase ranking module then orders them using an encoder-based ranker tuned on some domain-specific supervised signal. This simple generation framework presents two main issues: 1) the token space is limited by token adjacency and token presence in the item's text; 2) the keyphrase should also be in the universe of queries that buyers are searching for; which this simple generation model does not ensure, as described in \ref{sss:key_target}.\footnote{keyBERT can also use LLMs as generators, but their time complexity is substantial.} 

Table~\ref{tab:various_frameworks} shows how the various SOTA frameworks for keyphrase recommendation perform with the challenges mentioned previously. fastText, Graphite and SL-emb are all XMC tagging models deployed at eBay for seller-side keyphrase recommendations and are used in this study for comparison along with the rule-based models RE and SL-query.

\begin{table}[h]
\centering
\setlength\extrarowheight{2pt}
\begin{tabular}{c|ccc}
\toprule
\multirow{2}{*}{\textbf{Criteria}}                                                                & \multicolumn{3}{c}{\textbf{Frameworks}}                                                                                \\ \cmidrule{2-4} 
           & \multicolumn{1}{l}{\textbf{XMC-tagging}} & \multicolumn{1}{l}{\textbf{OOV}} & \multicolumn{1}{l}{\textbf{GraphEx}} \\ \midrule
\begin{tabular}[c]{@{}c@{}}Feasible daily batch\\ or real-time\\ prediction latency?\end{tabular} & \checkmark                &                                  & \checkmark            \\ \cmidrule{1-1}
\begin{tabular}[c]{@{}c@{}}Click data debiasing ?\end{tabular}                                  &                                          &                                  & \checkmark            \\ \cmidrule{1-1}
\begin{tabular}[c]{@{}c@{}}Susceptible to RE\\ De-duplication ?\end{tabular}                      &                                          & \checkmark        & \checkmark            \\ \cmidrule{1-1}
\begin{tabular}[c]{@{}c@{}}100\% targeting\\ in vocabulary\\ keyphrases ?\end{tabular}            & \checkmark                &                                  & \checkmark            \\ \cmidrule{1-1}
\begin{tabular}[c]{@{}c@{}}Focus on popular \\ keyphrases?\end{tabular}                          &                                          &                                  & \checkmark            \\ \bottomrule
\end{tabular}
\caption{Table showing the comparative analysis of the capabilities of the various types of frameworks for keyphrase recommendation.}
\label{tab:various_frameworks}
\vspace{-4mm}
\end{table}


\section{GraphEx Model}
\label{s:chap4_graphex}
We first formulate the keyphrase recommendation problem and then briefly go through the data set curation process. Next, we describe the notations, then the \textit{Construction} of the graph which is the training part of GraphEx and the \textit{Inference} method for obtaining the predictions.

\subsection{Problem Formulation}
\label{ss:problem_form}
For efficiently solving the recommendation problem, we use the formulation of a \textit{permutation} problem that permutes the title strings to match a given set of keyphrases. Let us consider a title string with $l$ words in it. The goal is to generate permutations of different lengths from the $l$ words. Now, given a list of predefined keyphrases, the possible permutations of the title string are constrained to match the keyphrases. Therefore, each permutation can exactly match a keyphrase or be part of some keyphrases, but if a title token is not part of any keyphrase then it is ignored. Thus, it does not limit the permutations to token adjacency or token presence in the item's text.  A naive brute force method is to generate all possible permutations of the $l$ words which will take $O(l!)$ time. Each keyphrase can be validated using hashing and string comparisons (each word can be an integer) and thus can take overall $O(l\times l!)$ time. This is infeasible to perform in real-time with limited amount of resources. 

\subsection{Dataset Curation}
\label{ss:chap4_datasets}
We aggregate our keyphrase datasets and their category associations from the search logs generated during buyer sessions on \textit{eBay.com}. The keyphrases that buyers input during the search sessions are curated based on certain criteria, which we discuss here. The categorization of items are in form of a tree. The root level category is called as meta category (top level category) which branches down to the \textit{Leaf Category}. eBay's search engine \textit{Cassini} shows a sufficient number of items (\textit{Recall Count}) for each input query in its search results. Cassini determines the leaf category of the keyphrase and it is the same as the top-ranked item's leaf category (lowest-level product categorization). Note that, as mentioned in Section~\ref{ss:chap4_scope_cont} the item-keyphrase click-based associations are not curated into our datasets, thus avoiding related biases and addressing Challenge~\ref{sss:click_biases}.

We restrict the number of curated keyphrases by only considering those that are heavily searched by the buyers (head keyphrases). This helps address Challenge~\ref{sss:bud_reco} by enabling GraphEx's final recommendations to have higher realizations adhering to budget constraints. The number of times a keyphrase is queried is termed as (\textit{Search Count}). The search count is used to explicitly control the proportions of head keyphrases in our dataset. The absolute values of Recall Count and Search Count are not essential in fact, an ranking (integer value) also works. All the unique keyphrases are aggregated for each meta category and are grouped for each leaf category within the meta category. Each keyphrase is associated with a Search Count and a Recall Count. Note that a keyphrase can be duplicated across different Leaf Categories. The curated dataset is used to construct GraphEx's graph (described later) from keyphrases that buyers actively search tackling Challenge~\ref{sss:key_target}.


\subsection{Terms and Notations}
\label{ss:chap4_notation_formulations}
We consider a set of unique keyphrases termed as $Q={k_1,k_2,...,k_K}$. Each keyphrase $k_i$ can be considered as a set of words ${w_1,w_2,...,w_l}$, where $w_*$ are tokenized\footnote{\label{fn:tokenization}The tokenization scheme can be anything as long as string comparison functions are well-defined and consistent for that scheme. By default we consider space-delimited tokenization.} from the keyphrase string $k_i$. Each $k_i$ is further associated with a Leaf category $l$, Recall Count or Rank $R$ and Search Count or Rank $S$. Given a test item's title $T$, the goal is to recommend a subset of keyphrases from $Q$ that are relevant to $T$. We can consider the title as a string with tokenized\footref{fn:tokenization} words $T={w_1,w_2,...,w_t}$ similar to a keyphrase, but titles are generally longer than the keyphrases. We denote a graph $G(V,E)$ where $V$ is the set of vertices and $E$ is the set of edges. Each edge $e\in E$ is denoted by a pair of vertices $e=(v_1,v_2)$ indicating a connection between the vertex pair. In a \textit{Bipartite Graph}, the set of vertices $V$ are divided into a pair of disjoint subsets $V=X\bigcup Y$. Each vertex in the same subset ($X$ or $Y$) isn't connected by an edge and only vertices in different subsets can be connected by an edge. We define the function \textit{Deduplicate and Count} or $DC(\cdot)$ which, given a list of elements, counts the occurrences of each unique element in the list. It outputs a list of tuples of the form $(element,count)$ for each unique element in the list.

\subsection{Construction Phase}
\label{ss:chap4_graphex_construction}
In this phase, the method relates the words in the keyphrases to the keyphrases themselves by mapping the relation using Bipartite Graphs. For a particular metacategory, the model constructs a series of Bipartite Graphs $G_l(V,E)$ one for each leaf category $l$ from only those keyphrases $Q_l$ that belong to the same leaf category. For each graph $G_l(V,E)$, the two subsets $X$ and $Y$ of the vertex set $V$ are constructed as follows: All the unique words in the keyphrases are considered as the set $X$, while the unique keyphrases are considered as $Y$. Each unique word and unique keyphrase is represented as non-negative integers, to avoid string comparison and manipulation costs. Mathematically, $X=\bigcup_{\forall w\in k_i,\forall k_i\in Q_l} \{w\}$ and $Y=Q_l$. An edge $e=(x,y)$ in set $E$, is permitted from vertex $x\in X$ to vertex $y\in Y$ when $x\subset y$, indicating an edge from a word to the keyphrase that it is a part of. Such edge relations are created for all the Bipartite Graphs using the unique words in all the unique keyphrases within each leaf category.

\begin{figure}[h]
\centering

\begin{subfigure}{8cm}
\centering
\includegraphics[width=\linewidth]{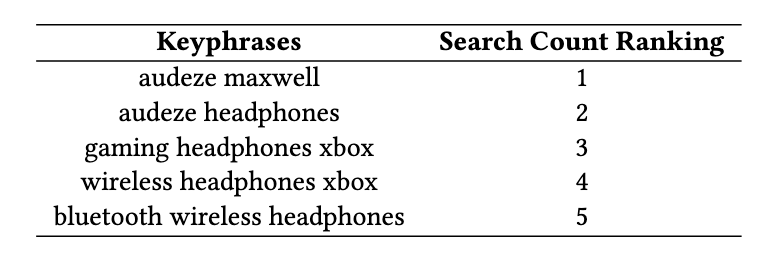}
  \caption{Illustrated Training Data}
  \label{fig:chap4_illus_data}
\end{subfigure}\qquad

\begin{subfigure}{8cm}
\vspace{4mm}
\centering
\includegraphics[width=0.8\linewidth]{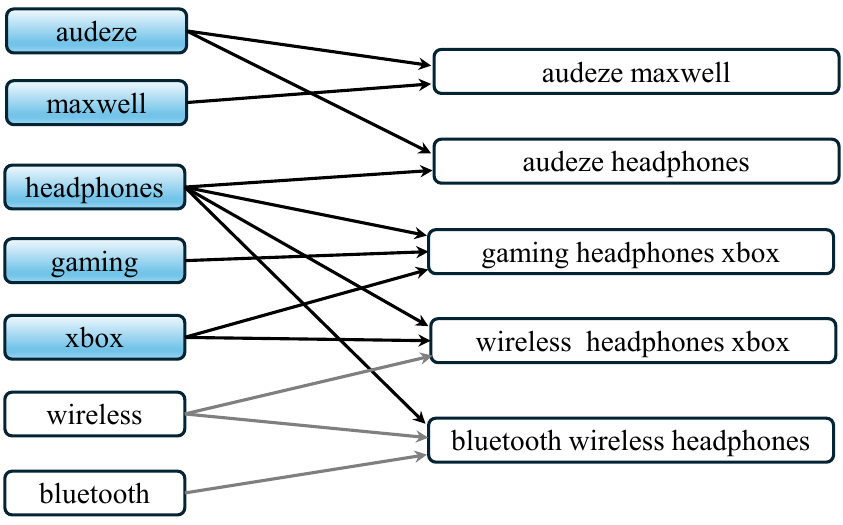}
\caption{Bipartite Graph derived from Illustrated Data}
\label{fig:chap4_illus_graph}
\end{subfigure}
\caption{Illustration of GraphEx's construction phase. (i) a set of keyphrases with their search volume rank, (ii) shows the bipartite graph constructed from the set in (i).}
\label{fig:chap4_graphex_const_eg}
\end{figure}

An example of a constructed Bipartite Graph is shown in Figure~\ref{fig:chap4_graphex_const_eg}. Each vertically stacked vertex belong to the same subset. The left set of vertices are the words/tokens and the right set are the keyphrases. The vertices are shown as strings here for presentation. Each tokenized\footref{fn:tokenization} word is connected to the keyphrase that it is a part of. The graph is stored in \textit{Compressed Sparse Row} (CSR) format, which occupies the least amount of space. Each word/token can be accessed in constant time whereas the adjacencies of a word can be traversed in $O(d)$ where $d$ is the degree of the word or the number of keyphrases that contain that word. A map type data-structure is used to associate the leaf category ID to the CSR structure for each graph. 

\begin{algorithm}[h]
\begin{algorithmic}[1]
\Require Graph $G_l(V,E)$, test item $T$ and each label's Search and Recall count
\Ensure List of lists ($C_R$) with labels and their attributes
\Function{Enumeration}{$G_l$,$T$}
\State $C_L,C_R \gets [\ ]$ \Comment{Lists of labels and results resp.}
\For {$w$ in $T$}
    \For {$(w,l)$ in $E\in G_l$}
        \State $C_L\gets C_L+l$
    \EndFor
\EndFor
\State $C_L\gets DC(C_L)$
\For{$(l,c)$ in $C_L$}
   \State{$C_R\gets C_R+(l,LTA(T,l,c),S(l),R(l))$}
\EndFor
\State \textbf{return} $C_R$
\EndFunction
\end{algorithmic}
\caption{GraphEx's Inference}
\label{alg:chap4_graphex_enum}
\end{algorithm}
\setlength{\textfloatsep}{0.1cm}
\setlength{\floatsep}{0.1cm}

The keyphrase's Recall and Search Count\footnote{Defined in section~\ref{ss:chap4_notation_formulations}} are stored in separate arrays. So given a keyphrase ID $l$, $R(l)$  and $S(l)$ will directly index into the arrays and return the values taking unit time. The space occupied by each leaf category graph depends linearly on the number of unique words and edges, as CSR structure occupies $|X|+|E|$ space. The count of edges $|E|$ depends on the sum total of occurrence of each word in the keyphrases/labels which is difficult to generalize and depends on the datasets. Separate graphs for each leaf category help in recommending more relevant keyphrases which becomes more clear in the next section.


\subsection{Inference Phase}
\label{ss:chap4_graphex_inference}
Given a test item $T$ and a leaf category $l$ with the tokenized words in the title as $T={w_1,w_2,...,w_t}$, the goal is to extract a list of keyphrases in decreasing order of relevance to the item. GraphEx's recommendation is based on permuting the words in the item's title as discussed in Section~\ref{s:chap4_introduction}. To enable this, the Inference Phase is divided into two steps: \textit{Enumeration} step that generates keyphrases from words of title and the \textit{Ranking} step that ranks the keyphrases in order of relevance to the item.

\vspace{2mm}
\subsubsection{Enumeration Step}
\vspace{1mm}
\label{sss:chap4_graphex_enum}
GraphEx first determines the Bipartite Graph $G_l(V,E)$ that corresponds to the leaf category $l$ of the input item. The corresponding graph $G_l$ can be obtained in $O(1)$ time if a hashing data-structure is used to map the leaf categories to the graphs defined in section~\ref{ss:chap4_graphex_construction}. 
The step first tokenizes the item's title into words and uses them as input along with the graph $G_l$ in the Algorithm~\ref{alg:chap4_graphex_enum}. Lines 3-5 of the algorithm map the tokenized words of $T$ using the bipartite graph $G_l$ to the labels/keyphrases. Let's look at an example to understand this process. Given an item \textit{``audeze maxwell gaming headphones for xbox''}, we highlight the corresponding words on the left in the illustrated Figure~\ref{fig:chap4_illus_graph}. The keyphrases ($l$) connected to the highlighted words are candidates for recommendation and are collected in $C_L$ in Algorithm~\ref{alg:chap4_graphex_enum}. Line 6 uses the $DC$ function to de-duplicate and count the redundancies in the candidate keyphrases. E.g. in Figure~\ref{fig:chap4_illus_graph} the keyphrase ``audeze maxwell'' is connected to two words ``audeze'' and ``maxwell'', whereas ``gaming headphones xbox'' is connected to 3 words. Hence, after the execution of line 6, it results in the duplication count of 2,2,3,2, and 1 in the given order for each of the keyphrases on the right side of the illustrated Figure~\ref{fig:chap4_illus_graph}. The count indicates the number of words in the keyphrase that are common with the item title $T$.

The next part of the Enumeration step generates a tuple corresponding to each label in $C_L$ using lines 7-9 in Algorithm~\ref{alg:chap4_graphex_enum}. We define the function \textit{Label Title Alignment} or $LTA$ that uses the common word count (or duplication count) $c=|T\cap l|$ between the title $T$ and the label $l$ as $LTA(l,c)=\frac{c}{|l|-c+1}$. The LTA ratio is the second element of the tuple or the first attribute of the label $l$. The two attributes are the Search $S(l)$ and Recall count $R(l)$ of the label. The tuples generated by this process are returned in $C_R$. The time complexity of this step primarily depends on lines 3-5 due to restriction on prediction count which we discuss later in Section~\ref{ss:chap4_impl_details}. The time complexity can be uncertain to determine due to the varying number of edges for each word. For simplification, we consider the average degree of each word as $d_{avg}=\frac{|E|}{|X|}$. Then asymptotically the time taken to gather the candidate labels for each word of the item title $T$ is $O(|T|.d_{avg})$. Modeling the problem as a Bipartite graph helps to efficiently permute all the words in the title $T$ while only generating permutations that are valid keyphrases. 

The Enumeration step aids in resolving the permutation problem laid out in Section~\ref{ss:problem_form}. Along with the curation process it retrieves keyphrases that are active and impactful, focusing on our targeting goal, as explained in Challenge~\ref{sss:key_target}.

\vspace{2mm}
\subsubsection{Ranking Step}
\vspace{1mm}
\label{sss:chap4_graphex_rank}
In this step, the candidate labels in $C_R$ are sorted in the non-increasing order of the first attribute or second tuple element \textit{LTA} and to break ties, $S(l)$ and subsequently $R(l)$ is used. While tie-breaking, those keyphrases are preferred that have higher search counts and lower recall counts. Higher search counts will have more clicks while lower recall count indicates the keyphrases have fewer items associated with them. So, when a keyphrase is input by a buyer, the search engine displays relatively fewer items, boosting click probability per item. The LTA function was designed to provide a higher score to those keyphrases that have less words in the label that aren't part of the title. Let's compare two keyphrases from the Figure~\ref{fig:chap4_illus_graph}, ``audeze maxwell'' and ``wireless headphones xbox'', both have 2 words in common with the sample title shown in section~\ref{sss:chap4_graphex_enum}. The first's LTA is $\frac{2}{1}$ and second's is $\frac{2}{2}$, thus ranking ``audeze maxwell'' higher. LTA minimizes the risk involved by preferring those keyphrases that have more complete information (or more matching words).

This results in more relevant keyphrases appearing at the top of recommendations, effectively addressing Challenge~\ref{sss:bud_reco}, especially when there is a limitation on the number of recommendations.

\subsection{Implementation Details}
\label{ss:chap4_impl_details}
GraphEx's inference algorithm and implementation have been designed for efficiency, aligning with our objectives for Challenge~\ref{sss:exec_perf_chal}. In this section, we examine the time complexity of the inference algorithm and explain how certain steps in its implementation are optimized to achieve real-time latency.

The edges of the bipartite graph of each leaf category are constructed as tuples, sorted and then de-duplicated based on their IDs which are finally stored in the CSR format.  The space complexity is linear in the number of edges for each graph given by $O(|X|\cdot d_{avg})$ where $X\in V$, which is the set of unique words in all the keyphrases for the leaf category. The words and the labels are represented as \textit{unsigned integers} to occupy minimal space and convert string comparisons to integer ones. Therefore, comparing two words or two labels takes $O(1)$ time. The construction phase does not involve any weight updates or hyper-parameter training, making it quite fast and efficient.


Generally, the leaf category ids within a meta category are unique. There are a large number of leaf categories ($> 100000$ in total). Models like fastText and Graphite were trained per meta category which are approximately 120. Other non-rule based models like SL were even more granular with item co-click based associations. GraphEx handles the subcategories internally and doesn’t need to train one model per leaf category, minimizing computational costs. Also, if the leaf categories are unique across various meta categories then only model will be required for the entire group. The leaf categories help GraphEx in recommending relevant keyphrases as generally the items and keyphrases in the leaf categories belong to the same product.

A drawback of directly using the Algorithm~\ref{alg:chap4_graphex_enum} is the large number of keyphrases that are generated in the initial $C_L$. This results in a poly-logarithm time complexity for line 6 in the algorithm. To circumvent this we used \textit{count} arrays to calculate the redundancies of each unit keyphrase. The space taken for the storage of $C_L$ and the count array is approximately $2 |Q_l|$. A predetermined number of keyphrases (10-20) are generated for a given test instance during the inference phase. So, after the counting in line 6, the number of unique keyphrases in $C_L$ is pruned based on this requirement. This is done by first grouping each keyphrase with similar counts, then restricting the number of groups so that the sum of group sizes is equal to the required number of predictions. Groups with larger number of keyphrase redundancy counts are preferred, and all keyphrases in the threshold group are included even if the group size exceeds the number of required predictions. Thus, the time complexity of the Enumeration step remains as $O(|T|.d_{avg})$. Although the sorting in the Ranking step seems expensive, the list length is always approximately a constant due to $|C_L|=|C_R|$. This is due to the restriction on prediction count as mentioned above thereby not contributing asymptotically to overall time complexity.


\subsection{Interpretability}
The applications in E-Commerce domain frequently require that a model be interpretable. This helps to comprehend the rational behind its predictions and decision process. In our use case, it is essential to trace where the words in the keyphrases come from.  Neural network models typically require converting input text into vectors, which often obscures the contribution of individual tokens to the decisions. Interpretability techniques such as LIME and SHAP offer post hoc explanations, treating a Deep Neural Network as a black-box. They also require much effort to figure out the contributions of each input feature.

Unlike black-box models, the GraphEx algorithm has 3 transparent phases: keyphrase curation, keyphrase mapping, and ranking. The data curation process gives perspective as to how the keyphrases in GraphEx's label set were curated. The keyphrase mapping phase details how GraphEx's candidate keyphrases were mapped from the keyphrases extracted from the item's title to GraphEx's candidate keyphrases. The ranking algorithm which then ranks the mapped candidates is transparent as well. It uses 
\textit{Label Title Alignment~(LTA)} outlined in Section~\ref{sss:chap4_graphex_enum} which is a token-based algorithm ensuring that the majority of tokens in the keyphrases match the title. This ensures that GraphEx's predicted keyphrases are explainable and interpretable.

\section{Experimentation and Results}
\label{s:cha4_experimentation_results}
We perform experiments on representative datasets from \textit{eBay} and compare the results of our model with the described models (Section~\ref{s:chap4_related_work}) in production at eBay. We first describe our experimental setup, the datasets we use and the models we compare in Section~\ref{ss:chap4_experimentation_details}.
Next, we describe our evaluation framework in Section~\ref{sss:chap4_metrics} on how we determine relevance and the metrics using relevance for performance comparison.
Then, we analyze the results of each model's performance in~\cref{ss:chap4_perf_results,ss:misc_results}, ablation studies in~\ref{ss:chap4_aba_studies} and the execution performance of each model in ~\ref{ss:chap4_exec_results}. Finally, we describe the deployment in production in Section~\ref{ss:deployment} and its impact in section~\ref{ss:impact}.

\begin{table}[h]
\centering
\begin{tabular}{llll}
\toprule
\textbf{MetaCat} & \textbf{\# Items} & \textbf{\# Keyphrases} & \textbf{\begin{tabular}[c]{@{}l@{}}\# GraphEx \\ Keyphrases\end{tabular}} \\
\midrule
CAT\_1           & 200 M             & 3.6 M                  & 115 K                                                                     \\
CAT\_2           & 14 M             & 0.83 M                 & 252 K                                                                     \\
CAT\_3           & 7 M            & 0.46 M                 & 47 K     \\
\bottomrule
\end{tabular}
\caption{Details of three representative meta categories of eBay.}
\label{tab:chap4_datasets}
\vspace{-4mm}
\end{table}

\begin{figure*}[t]
    \centering
         \includegraphics[width=\linewidth]{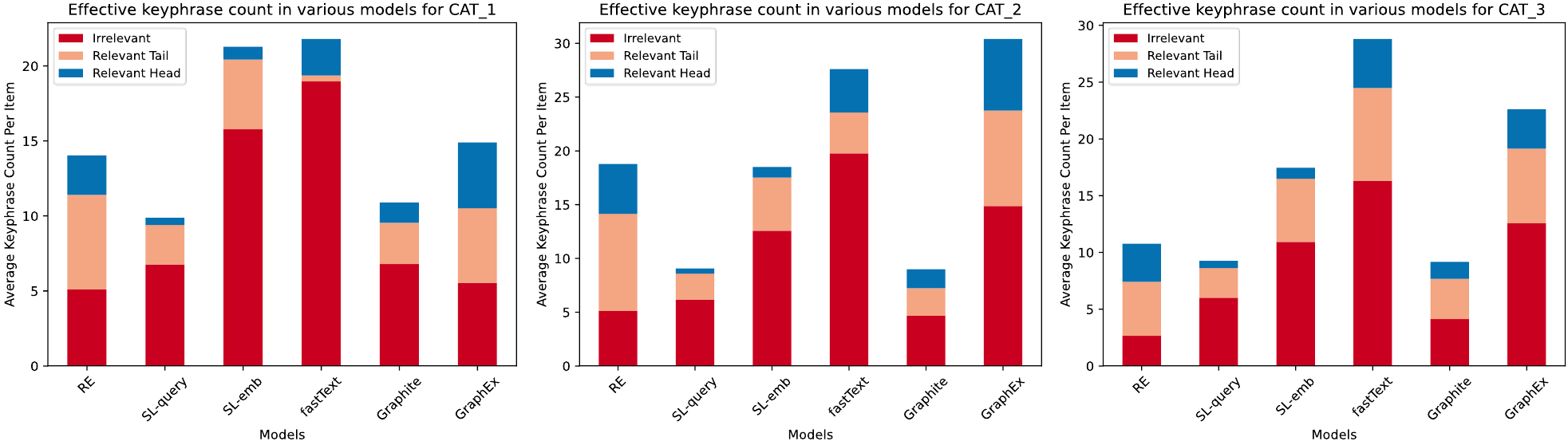}
      \caption{The average counts of relevant head/tail and irrelevant keyphrases per item are shown for each model.}
      \label{fig:chap4_graphex_abs_perf}
\vspace{-4mm}
\end{figure*}

\subsection{Setup and Datasets}
\label{sss:chap4_setup_datasets}
GraphEx is implemented for multi-core systems without requiring any GPUs. Its inference part is implemented on C++ ($\geq$ g++-9.3.0) using OpenMP threading with Python wrappers using \textit{pybind11}. The construction part is implemented in Python ($\geq$ 3.7); due to its lightweight approach since the construction does not require large resources and takes much less time. We used a system with 4 Intel Xeon Gold 6230R CPUs with 2 sockets each containing 20@2.10GHz cores, and 500 GB of RAM for the analysis. A single machine is utilized here for test standardization. GraphEx employs coarse-grained multithreading, assigning each input's inference to an individual thread. We launched 20 threads with compact pinning to occupy only a single socket sufficient for our dataset size.

\subsection{Experimentation Details}
\label{ss:chap4_experimentation_details}

We present findings on three product meta categories from eBay, each symbolizing a classification of large, medium, and small categories. The classification is determined by the count of items and the quantity of unique keyphrases within each meta-category. Table~\ref{tab:chap4_datasets} shows the anonymized categories and their details. Even though our methodology does not require knowledge of the items or their meta-data, the XMC models require them, and hence we show their numbers for perspective. Our data curation and analysis are limited to eBay, due to the absence of any publicly available keyphrase recommendation datasets from e-commerce advertisement platforms.

The data is collected from search logs for the duration of one year for XMC models and 6 months for GraphEx. For XMC, the item-keyphrase pairs are constrained based on their co-occurrence count, number of buyer clicks/purchases, etc. 
The curated unique keyphrase count shown in the third column of table~\ref{tab:chap4_datasets} contains both the head and tail keyphrases and is incorporated by XMC models. On the other hand, GraphEx's data curation for training, aggregates keyphrases without looking at any (clicked-based) association with the items. It restricts the keyphrases to contain a higher number of head and a lower number of tail keyphrases using the curation process described in Section~\ref{ss:chap4_datasets}. Generally, keyphrases that on an average weren't searched at least once per day were filtered for GraphEx\footnote{The constraint was eased for CAT\_3 due to a lack of enough keyphrases.}.

For testing, we sampled a set of 1000, 400 and 200 items from actively listed items on eBay.com for the categories CAT\_1, CAT\_2, and CAT\_3 respectively. We also computed the search count of each unique keyphrase by considering a 15 day duration different from the one year duration for the training set. This removes any bias that models have based on their training data. For each of the test items, all the models generate a variable number of keyphrases with a limit of 40. 


\subsection{Evaluation Framework}
\label{sss:chap4_metrics}
We describe here our evaluation framework that tackles the challenges in Section~\ref{sss:click_eval}. Summarizing the two major problems are 1) Lack of ground truths due to sparse data and MNAR-led biases in click data and 2) Model Convergence due to similar training data and evaluation. Ideally, metrics should compare the relevancy of the predictions to the input text without limiting the comparison to a set of predefined labels/keyphrases. However, it is difficult to determine the relevance of predictions without any prior labels or the absence of negative labels. Thus, directly using traditional metrics like precision and recall which require ground truths incorporates the biases described in Section~\ref{sss:click_eval}.

So, while previous research has used human judgement~\cite{beyondPosBias}, we use AI-generated evaluations to evaluate at scale for the variety of items at eBay.\footnote{The AI predictions were benchmarked against positive buyer judgement and achieved more than 90\% alignment, similar to how it was done in \cite{ashirbad-etal-2024}.} We generate prompts for \textit{Mixtral 8X7B}\footnote{We experimented with GPT4~\cite{openai2023chatgpt} model and the results are comparable. We use Mixtral because of costs and API rate limits.} \cite{jiang2024mixtralexperts} per item, which contains the item's title and a set of predicted keyphrases. The structure of the prompt is shown below. The response is ``yes'' or ``no'' for each keyphrase, indicating whether the keyphrase is relevant to the item or if it is irrelevant.

\begin{tcolorbox}
\noindent\texttt{\footnotesize{Below is an instruction that describes a task. Write a response that appropriately completes the request. \\
\\
\#\#\# Instruction: \\
Given an item with title: "\{\textrm{title}\}", determine whether the keyphrase: "\{\textrm{keyphrase}\}", is relevant for cpc targeting or not by giving ONLY yes or no answer: \\
\\
\#\#\# Response:
}}
\end{tcolorbox}

Keyphrases identified as pertinent to an item undergo filtering through a high \textit{Search Count} threshold, which is set at the 90\textsuperscript{th} percentile of search counts for all unique keyphrases in the category, ensuring 10\% exceed this limit. Those surpassing this threshold are labeled as \textit{Relevant Head Keyphrases}, whereas others are deemed \textit{Relevant Tail Keyphrases}.\footnote{This evaluation method is solely for offline analysis; the model's seller recommendations are not filtered this way due to LLM latency constraints.} The rationale is that keyphrases deemed irrelevant by AI will not perform effectively; buyers may frequently search them, but seldom click on the corresponding item.

We compare the models based on the effective (relevant and head) keyphrases that each model recommends. Figure~\ref{fig:chap4_graphex_abs_perf} shows the per-model number of keyphrases averaged over all items that are evaluated as relevant or irrelevant by AI, while also distinguishing the head and tail types in the relevant keyphrases. The x-axis shows all the models under comparison. The y-axis shows the average number of keyphrases per item that are irrelevant and relevant head/tail keyphrases, while summing up to the total predictions by each model.

\begin{table*}[t]
\centering
\begin{tabular}{c|ccc|ccc|ccc|ccc}
\toprule
\multirow{2}{*}{\textbf{Models}} & \multicolumn{3}{c|}{\textbf{\begin{tabular}[c]{@{}c@{}}RP\end{tabular}}} & \multicolumn{3}{c|}{\textbf{\begin{tabular}[c]{@{}c@{}}HP\end{tabular}}} & \multicolumn{3}{c|}{\textbf{\begin{tabular}[c]{@{}c@{}} RRR\end{tabular}}} & \multicolumn{3}{c}{\textbf{\begin{tabular}[c]{@{}l@{}}RHR\end{tabular}}} \\
\cmidrule{2-13}
& \textbf{CAT\_1}                  & \textbf{CAT\_2}                  & \textbf{CAT\_3}                 & \textbf{CAT\_1}                 & \textbf{CAT\_2}                & \textbf{CAT\_3}                & \textbf{CAT\_1}                                & \textbf{CAT\_2}                               & \textbf{CAT\_3}                               & \textbf{CAT\_1}                              & \textbf{CAT\_2}                              & \textbf{CAT\_3}                             \\
\midrule
fastText                          & 13.1\%                           & 28.4\%                           & 43.4\%                          & 11.3\%                          & 14.6\%                         & 14.9\%                         & 0.31                                          & 0.51                                         & 1.25                                         & 0.55                                        & 0.61                                        & 1.24                                       \\
SL-emb                           & 25.9\%                           & 32.1\%                           & 37.4\%                          & 3.99\%                          & 5.35\%                         & 5.56\%                         & 0.59                                          & 0.38                                         & 0.65                                         & 0.19                                        & 0.15                                        & 0.28                                       \\
SL-query                         & 31.6\%                           & 31.9\%                           & 35.2\%                          & 4.86\%                          & 5.41\%                         & 6.91\%                         & 0.33                                          & 0.19                                         & 0.33                                         & 0.11                                        & 0.07                                        & 0.19                                       \\
Graphite                         & 37.9\%                           & 48.1\%                           & 55.1\%                          & 12.5\%                          & 19.6\%                         & 16.2\%                         & 0.44                                          & 0.28                                         & 0.5                                          & 0.31                                        & 0.26                                        & 0.43                                       \\
RE                               & 63.7\%                           & 72.8\%                           & 75.5\%                          & 18.7\%                          & 24.7\%                         & 31.2\%                         & 0.95                                          & 0.88                                         & 0.81                                         & 0.59                                        & 0.69                                        & 0.97                                       \\
GraphEx                          & 56.4\%                           & 51.1\%                           & 44.4\%                          & 26.5\%                          & 21.9\%                         & 15.27\%                        & 1                                             & 1                                            & 1                                           & 1                                           & 1                                           & 1                                          \\ 
\bottomrule
\end{tabular}
\caption{Comparing all models on $\mathit{RP}$, $\mathit{HP}$, $\mathit{RRR}$ and $\mathit{RHR}$. $\mathit{RRR}$ and $\mathit{RHR}$ were computed w.r.t GraphEx.}
\label{tab:chap4_performance_ratios}
\vspace{-2mm}
\end{table*}

It is evident from Figure~\ref{fig:chap4_graphex_abs_perf} --- as the number of predictions generated by a model increases, the number of irrelevant predictions also tends to rise. Our evaluation metrics are as follows:

\begin{itemize}
    \item $\displaystyle\mathit{Relevant}\; {Proportion~(RP)=}\frac{\#\;\mathit{relevant}\;\mathit{predictions}}{\#\;\mathit{total}\;\mathit{predictions}}$
    \vspace{2mm}
    \item $\displaystyle\mathit{Head}\; {Proportion~(HP)=}\frac{\#\;\mathit{head}\;\mathit{predictions}}{\#\;\mathit{total}\;\mathit{predictions}}$
    \vspace{2mm}
    \item $\displaystyle\mathit{Relative}\;\mathit{Relevant}\;\mathit{Ratio}\; \mathit{(RRR)}=\\
    \phantom{{}=1+2+3+4+5}\frac{\#\;\mathit{relevant}\;\mathit{model1}\;\mathit{predictions}}{\#\;\mathit{relevant}\;\mathit{model2}\;\mathit{predictions}}$
    \vspace{2mm}
    \item $\displaystyle\mathit{Relative}\;\mathit{Head}\;\mathit{Ratio}\; \mathit{(RHR)}=\\
    \phantom{{}=1+2+3+4+5+6}
    \frac{\#\;\mathit{head}\;\mathit{model1}\;\mathit{predictions}}{\#\;\mathit{head}\;\mathit{model2}\;\mathit{predictions}}$
\end{itemize}
\vspace{2mm}
Due to the varying number of predictions by each model, we use one set of metrics to compare the relevant and head keyphrases within each model (RP and HP) and another set to compare between different models (RRR and RHR). By disassociating from traditional metrics we avoid relying on ground truth biases thus addressing Challenge~\ref{sss:click_eval}.

\subsection{Performance Results}
\label{ss:chap4_perf_results}
We use the metrics defined in the previous section to compare each model's predictions.

\vspace{2mm}
\subsubsection{Performance Comparison}
\vspace{1mm}

Table~\ref{tab:chap4_performance_ratios} demonstrates the evaluations using both sets of metrics on relevant and head keyphrases. The metrics RRR and RHR are calculated using the GraphEx's predictions as the denominator (\textit{model2}). It is important to note that each set of metrics alone do not offer a comprehensive view. Depending on the variation in the total predictions between the two models, RP and HP tend to favor the model with fewer predictions, while RRR and RHR favor \textit{model1} if it has a higher count. We do not show absolute numbers due to the proprietary nature of data and the models.

For clarity, we first discuss the models that have a much large number of predictions, as seen in Figure~\ref{fig:chap4_graphex_abs_perf} which are SL-emb and fastText. For Table~\ref{tab:chap4_performance_ratios}, fastText and SL-emb have lower RP and HP (columns $2^{nd}$ and $3^{rd}$) than GraphEx due to their large prediction count. However, we can also see that in RRR and RHR (columns $4^{th}$ and $5^{th}$), GraphEx outperforms fastText (except CAT\_3) and SL-emb in all categories. Thus, GraphEx has a lower percentage of irrelevant keyphrases and a higher count of relevant and head keyphrases. CAT\_3 is a small metacategory with fewer items and lower buyer interaction, leading to fewer keyphrases. Therefore, creating effective keyphrases for GraphEx becomes difficult and necessitates tailored curation. 

The models that have a comparatively smaller total count of predictions are RE, SL-query, and Graphite. In Table~\ref{tab:chap4_performance_ratios}, it is evident that these models (except SL-query) have a higher RP compared to GraphEx. This is attributable to their lower number of predictions, which skews the proportions. However, excluding RE and Graphite, all models exhibit HP significantly smaller than that of GraphEx. Additionally, all the models have much smaller RRR and RHR. Although Graphite has a slightly higher HP for CAT\_3, its RRR and RHR are still lower than that of GraphEx for all categories. Consequently, models are unlikely to achieve substantial clicks like GraphEx due to the fewer head keyphrases.

The model RE is a simple retrieval technique, based on recalling the ground truth (item-query combinations with associated buyer activity) with a minimum amount of buyer activity in a short lookback period. The results of RE as seen in Table~\ref{tab:chap4_performance_ratios} are mixed, with lower HP in CAT\_1, while $2.8\%$ and $15.9\%$ more HP than GraphEx in CAT\_2 and CAT\_3 respectively. The RRR and RHR of RE is always lower than GraphEx. Albeit its simple nature, RE is a 100\% recall recommender that reflects the ground truth in terms of actual buyer-engagement.

\vspace{2mm}
\subsubsection{Diversity Comparison}
\vspace{1mm}

While we covered the two aspects of comparison, lower irrelevant and higher head keyphrase counts; diversity is another aspect that determines whether the effective keyphrases generated by a model will bring substantial incremental impact. In complicated systems like e-commerce where we have multiple retrieval sources the incremental impact of any model is dictated by the amount of clicks/sales/revenue brought about by \textit{only the keyphrases that are unique to the model, i.e., not present in any other retrieval sources}. A diverse set of keyphrases is beneficial as it typically results in more engagement, especially if the keyphrases are relevant and are head keyphrases. Based on feedback from sellers, they prefer a balance: they dislike having either too many or too few keyphrases. \textit{Thus the ultimate goal is to predict a reasonable number of total keyphrases with a higher proportion of relevant keyphrases and a diverse set of head keyphrases.}

\begin{figure}[t]
    \centering
    \includegraphics[width=0.65\linewidth]{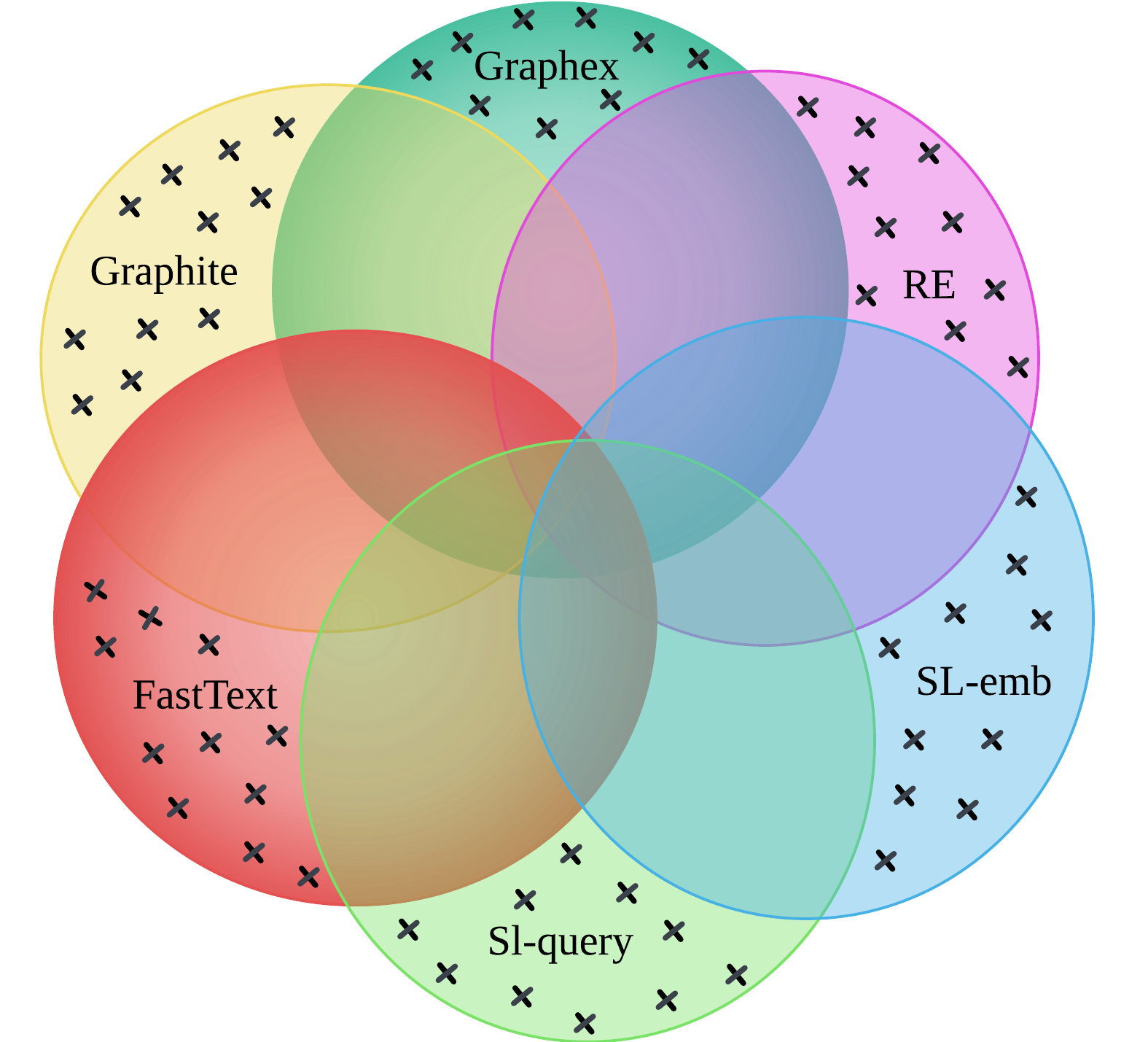}
    \caption{Venn diagram illustrating various recall sources and their associated keyphrases. The intersecting areas show keyphrases shared by multiple sources, whereas the crossed-out, non-overlapping sections represent keyphrases unique to each source, highlighting their distinct contributions.}
    \label{fig:diversity_illus}
\end{figure}

Therefore, the final metric to compare with other models is the exclusive diversity of GraphEx's predictions. For this evaluation, as illustrated in Figure~\ref{fig:diversity_illus}, we first separate out the unique head keyphrases recommended by each model that are relevant to the item. Table~\ref{tab:chap4_graphex_diversity} shows the relative amount of relevant exclusive head keyphrases of each model to the relevant exclusive head keyphrases of GraphEx (averaged per item). It is evident that GraphEx recommends the highest amount of diverse relevant head keyphrases in contrast to any other model. Since this evaluation does not depend on ground truths, we can better assess the number of unique keyphrases that each model recommends, thus also assessing model convergence, which is a key challenge in \ref{sss:click_eval}.


\begin{table}[h]
\centering
\begin{tabular}{c|ccc}
\toprule
\textbf{Models}   & \textbf{CAT\_1} & \textbf{CAT\_2} & \textbf{CAT\_3} \\
\midrule
fastText & 1.88x  & 2.36x  & 1.03x  \\
SL-emb   & 5.07x  & 5.63x  & 1.87x  \\
SL-query & 8.72x  & 12.2x  & 3.82x  \\
Graphite & 3.06x  & 3.26x  & 1.44x  \\
RE       & 1.57x  & 1.57x  & 1.11x \\
\bottomrule
\end{tabular}
\caption{Relative amount of relevant diverse (exclusive) Head keyphrases in GraphEx in comparison to other models. Each model's predictions are standalone.}
\label{tab:chap4_graphex_diversity}
\end{table}
\vspace{-2mm}

\subsection{Miscellaneous Results}
\label{ss:misc_results}
For sake of completeness, we compare the precision and recall scores of all the models. As discussed earlier, RE uses the click-based keyphrases for each item, hence we use RE's recommended keyphrases as the ground truths to compare other models' predictions with it. Table~\ref{tab:rel_prec_recall} shows the precision and recall scores of each model relative to GraphEx.

\begin{table}[h]
\centering
\begin{tabular}{@{}l|llll@{}}
\toprule
 \textbf{Metrics}  & \textbf{fastText} & \textbf{Graphite} & \textbf{Sl-emb} & \textbf{SL-query} \\ \midrule
Precision & 1.08     & 1.84     & 0.87   & 0.95     \\
Recall    & 1.09     & 1.62     & 4.01   & 3.43     \\ \bottomrule
\end{tabular}
\caption{Relative Precision and Recall numbers of other models with respect to Graphex. }
\label{tab:rel_prec_recall}
\end{table}

The numbers reflect that GraphEx has the lowest recall and low precision scores, indicating minimal ground truth retrieval capability. This works in its favor as the de-duplication with RE (recall) is minimal. This is further reinforced in the impact where GraphEx's incremental impact is superior to the other models.

\subsection{Abalation Studies}
\label{ss:chap4_aba_studies}
\vspace{2mm}
\subsubsection{Alignment functions}
\vspace{1mm}
We experimented with various alignment functions other than LTA, such as Word Match Ratio (WMR) used by Graphite~\cite{ashirbad-etal-2024} and the Jaccard coefficient (JAC). In terms of the notations used for LTA in section~\ref{sss:chap4_graphex_enum}, we re-define as $WMR=\frac{c}{|l|}$ and $JAC=\frac{c}{|l|+|T|-c}$. Table~\ref{tab:chap4_graphex_wmrnjac} compares the results when we use the three alignment functions in the GraphEx algorithm and generate recommendations for the sample of items in the previous section.

It might seem like JAC is the same as LTA, since the only differentiation is the term $|T|$ in the denominator of JAC. This term is constant when re-ranking the candidate keyphrases for a single input title. Hence both LTA and JAC should behave similarly, which is somewhat true from table~\ref{tab:chap4_graphex_wmrnjac}. However, there is still a difference because $|T|$ is much larger than $|l|$ which generally makes the numerator differentiate among the keyphrases. Let's take an example title with 10 tokens (A-J) which we will compare with two keyphrases ``A B C" and ``A B C D E", LTA ranks higher the former ($\frac{3}{1} > \frac{4}{2}$) while JAC ranks the latter higher ($\frac{3}{10} < \frac{4}{10}$). The token ``E" is risky, as it can be incorrect or change the product entirely; this risk is minimized by using LTA which penalizes such mismatches. Large-scale experiments with over 110 million items have shown at least 5\% points difference in RP between both the functions.

\begin{table}[h]
\centering
\begin{tabular}{@{}c|ccc@{}}
\toprule
\multirow{2}{*}{\textbf{Category}} & \multicolumn{3}{c}{\textbf{RP} (\%)} \\ \cmidrule(l){2-4} 
                          & \textbf{WMR }    & \textbf{JAC}     & \textbf{LTA}     \\ \midrule
CAT\_1                    & 33.6    & 44.5    & 45.8    \\
CAT\_2                    & 40.8    & 40.8    & 40.8    \\
CAT\_3                    & 42.6    & 55.0    & 56.0    \\ \bottomrule
\end{tabular}%
\caption{Relevant proportion (RP) of Word Match Ratio (WMR), Jaccard Coefficient (JAC) and Label Title Alignment (LTA) when used in GraphEx algorithm for the 3 categories.}
\label{tab:chap4_graphex_wmrnjac}
\end{table}

Additional alignment techniques such as semantic matching were considered but avoided due to latency constraints. We experimented with subword matching to improve our token matching function; this increased the inference latency without too much improvement in performance. We used a proprietary stemming function for words to increase the reach of token matches.

\vspace{2mm}
\subsubsection{Data Curation Effects}
\vspace{1mm}
\label{sss:graphex_ablation_datacuration}
A critical component of GraphEx's training involves the process of data curation. We find that the \textit{Search Count} defined in Section~\ref{ss:chap4_datasets} is crucial for predicting relevant as well as head keyphrases. A low Search Count of 1 inculcates many bogus user queries and hence needs a much higher threshold. An ideal threshold would be keyphrases that are queried at least once daily, which equates to 180 over a span of 6 months. However, as indicated in Table~\ref{tab:chap4_datasets}, this threshold results in a reduced number of unique keyphrases, necessitating a relaxation of the limit.

To comprehend the influence on recommendations, we evaluated two GraphEx models constructed with search counts of 90 and 180, respectively. A random subset of 1000 items from CAT\_1 was utilized for testing. Approximately $20.1\%$ of the items had identical recommendation sets from both models. For the remaining $80\%$ of items, $20\%$ had similar relevant keyphrases and $7.2\%$ had the same relevant head keyphrases. For the remaining disparate recommendations (about 60\%), the proportions of relevant and head keyphrases for the Search Count thresholds of 90 and 180 are presented in Table~\ref{tab:chap4_sc_threshold}. The benefit obtained with head keyphrases at the 180 search count surpasses the benefit obtained for relevant keyphrases at the same count when compared to a search count of 90. 

\begin{table}[h]
\centering
\begin{tabular}{@{}c|cc@{}}
\toprule
\begin{tabular}[c]{@{}c@{}}\textbf{Search Count} \\ \textbf{Threshold}\end{tabular} &
  \begin{tabular}[c]{@{}c@{}}\textbf{\% Relevant} \\ 
\textbf{Keyphrases}\end{tabular} &
  \begin{tabular}[c]{@{}c@{}}\textbf{\% Relevant}\\ \textbf{Head Keyphrases}\end{tabular} \\ \midrule
90 &
  12.2 &
  0.43 \\
180 &
  10.1 &
  5.64 \\ \bottomrule
\end{tabular}
\caption{Percentage of relevant and head keyphrases (exclusive) for training curated with different Search Count thresholds.}
\label{tab:chap4_sc_threshold}
\vspace{-4mm}
\end{table}

\subsection{Execution Results}
\label{ss:chap4_exec_results}

It is important for the models to attain the real-time recommendation and model refresh goals as described in section~\ref{sss:exec_perf_chal}. We compare only the XMC models with GraphEx as the \textit{REs} and SLs (except SL-emb) are simple retrieval techniques implemented in the Spark/Hadoop ecosystem while model inferencing is more complex technique. We examine the models based on Inference Latencies, Model Sizes, and Training times. \footnote{The inference phases of SL-emb are intricate, as embedding creation is processed on the GPU while ANN operations occur on the CPU, complicating latency comparisons with alternative models.}

\begin{figure}[h]    
\centering
    \begin{subfigure}[b]{\linewidth}
    \centering
    \includegraphics[scale=0.3]{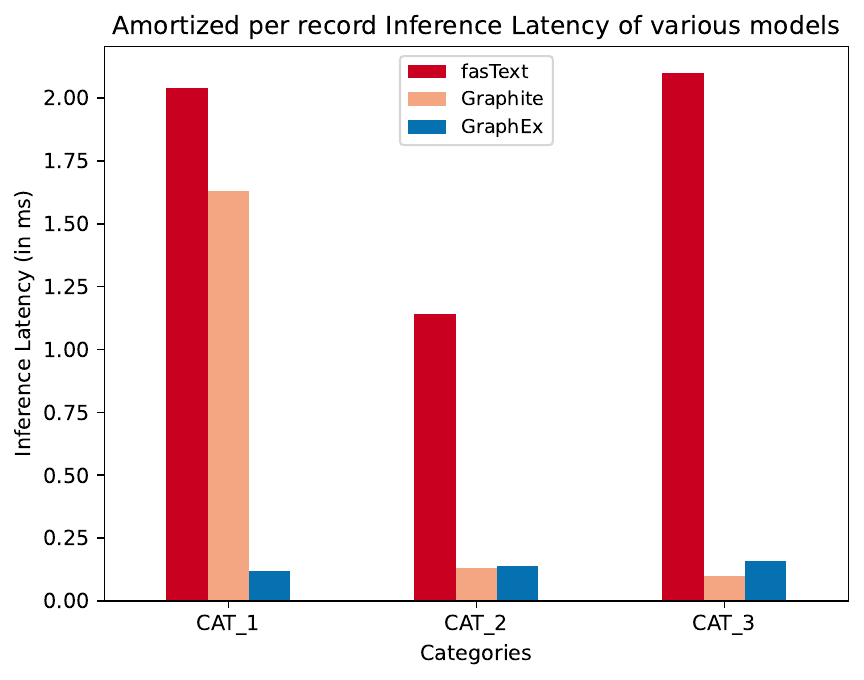} 
    \caption{Inference latency in milliseconds.}
    \label{fig:graphex_execution_inference}
    \end{subfigure}
    \begin{subfigure}[b]{\linewidth}
    \centering
        \includegraphics[scale=0.3]{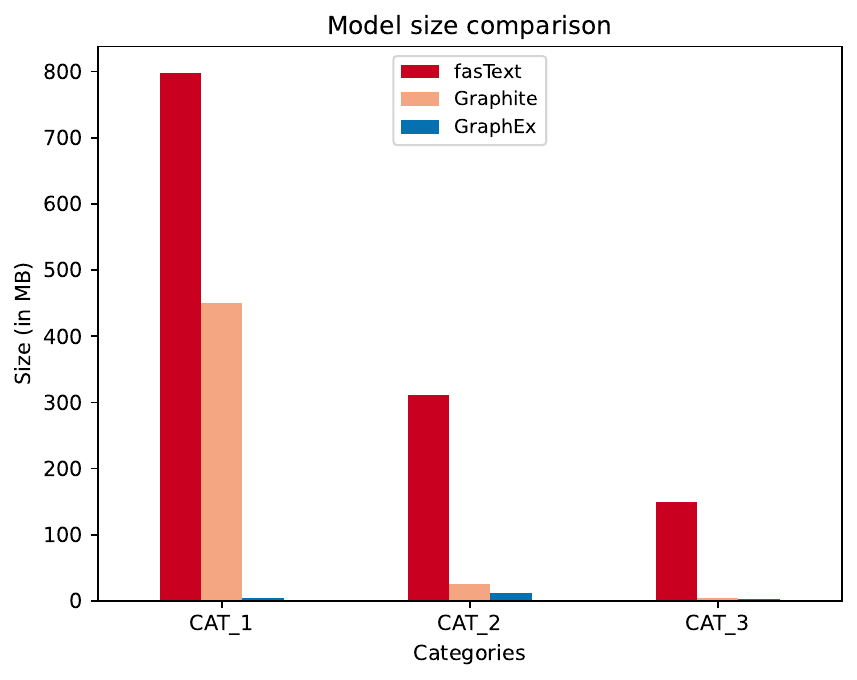}
    \caption{Model sizes in megabytes.}
    \label{fig:graphex_execution_modelsize}
    \end{subfigure}
    \caption{Execution performances of fastText, Graphite, and GraphEx.}
    \label{fig:chap4_graphex_execution}
\end{figure}

For near real-time recommendation, the Inference Latency of a single input should be in milliseconds. Figure~\ref{fig:graphex_execution_inference} compares the per input inference latency of the XMC models and GraphEx. The latencies for each model are computed by amortizing the time taken for prediction over the entire test set. We can see that all the models are within the required limit of 10 ms, but fastText takes more time for a prediction. Graphite and GraphEx's latencies are comparable for the smaller categories (CAT\_2 and CAT\_3). The performance of GraphEx is superior, attaining up to $17\times$ and up to $13\times$ more speed up in contrast to fastText and Graphite on CAT\_1. If we infer 20 million items in CAT\_1, GraphEx will result in energy savings of 11 hours and 8.5 hours with respect to fastText and Graphite, respectively.


Figure~\ref{fig:graphex_execution_modelsize} evaluates model storage demands. fastText consumes considerably more memory due to its weight matrix and word embeddings, despite a size reduction later to boost production accuracy. Graphite's storage demand is sizeable for CAT\_1 but aligns with GraphEx in other cases. GraphEx maintains minimal storage requirements, even with multiple leaf category graphs. FastText training surpasses $4$ hours, extending to several days for larger groups, encompassing extensive epochs and autotuning. Graphite's graph construction spans $1-6$ minutes, whereas GraphEx completes in under $1$ minute, aided by focused data training and effective model construction execution.

\begin{figure*}[t]
    \centering

\includegraphics[width=0.6\linewidth]{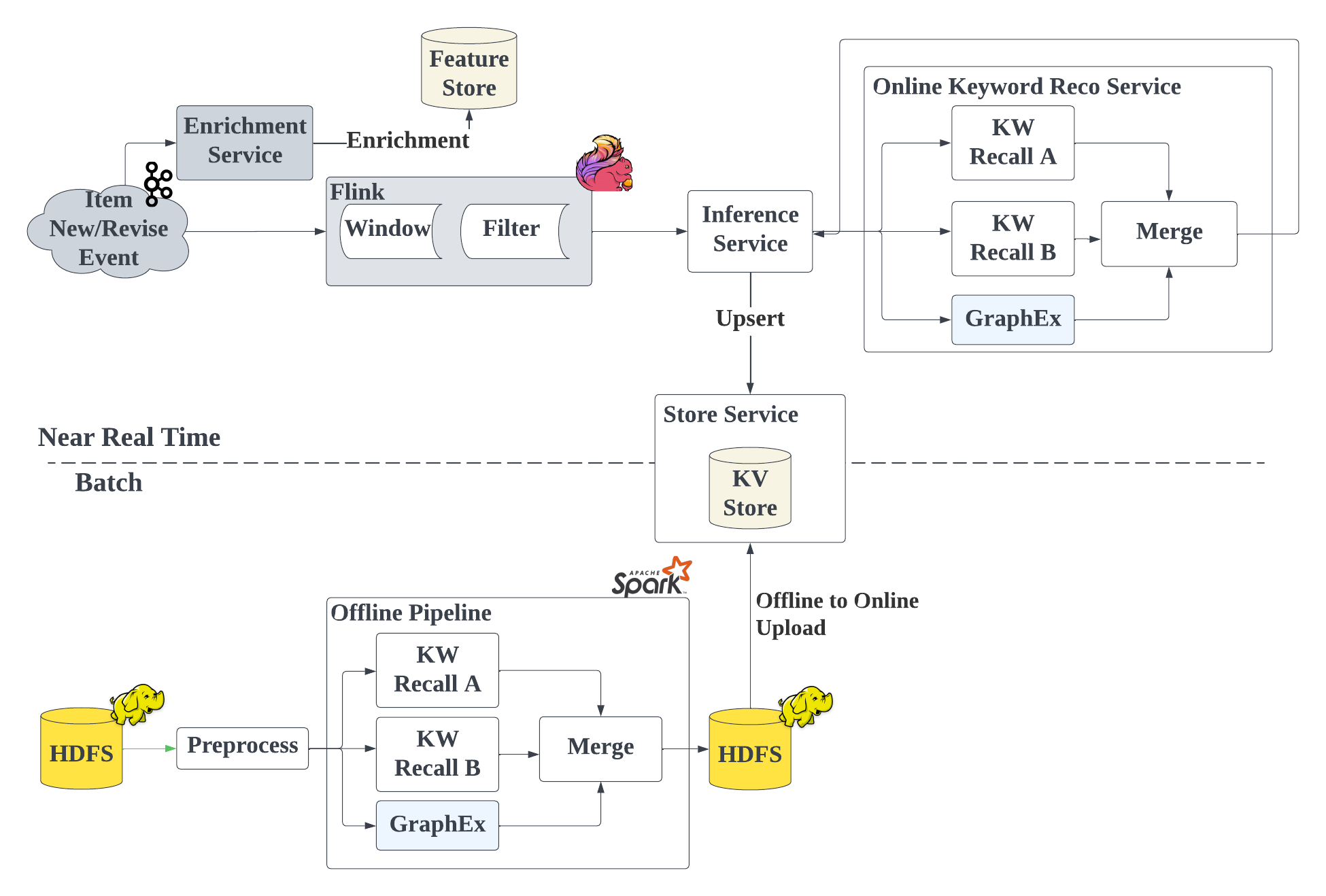}  
    \caption{GraphEx Batch/NRT Serving Architecture}
    \label{fig:serving_arch}
\end{figure*} 

\subsection{Production Engineering Architecture}
\label{ss:deployment}

In this section, we describe the engineering architecture used to serve GraphEx keyphrases to our sellers for their inventories in one of eBay's major sites. There are two components for recommendation \textit{Batch} and \textit{Near Real-Time (NRT)} Inference. Batch inference primarily serves items with a delay, whereas NRT serves items on an urgent basis, such as items newly created or revised by sellers. The batch inference is done in two parts: 1) for all items in eBay, and 2) daily differential, i.e. the difference of all new items created/revised and then merged with the old existing items. The NRT inference is done using Python code hosted by eBay's internal ML inference service Darwin. Darwin is then called by eBay's recommendation service, triggered by the event of new item creation or revision, behind a Flink processing window and feature enrichment. Note that GraphEx serves as one of the keyphrase recommendation sources in the whole Batch/NRT framework.

GraphEx's batch inference leverages eBay's Krylov machine learning platform\cite{kr19} and is executed on a single node featuring 70 cores and 900 GB RAM. The inference task for 200 million items for one site completes in merely 1.5 hours, vastly surpassing the performance of fastText and Graphite, which require 1.75 and 1.5 days, respectively.\footnote{In production, GraphEx can operate in a distributed manner, adapting to resource availability and workload. It achieves parallelization by replicating the constructed graph across all machines and evenly distributing the inference load for multiple sites.} An additional Spark batch process utilizes Hadoop to integrate these sources into NuKV, a Key-Value store accessed via eBay's inference API, subsequently serving sellers on the platform. This system, depicted in Figure ~\ref{fig:serving_arch}, scales efficiently to handle billions of items and keyphrases. GraphEx's algorithm is limited by the label space it uses for training but remains cost-effective like Graphite, requiring only minutes to train even large category models, enabling \textit{daily model refresh}. This allows adaptation to new keyphrases arising daily, significantly improving upon fastText, which demands over a day for training large categories and updates monthly.

\subsection{Impact}
\label{ss:impact}

GraphEx was deployed for the sellers of a particular site on eBay to replace Graphite keyphrases. After its release, a differential pre-post analysis was done to gauge the impact of GraphEx keyphrases in comparison to Graphite which it replaces. The differential analysis also involved measuring the impact of all keyphrases generated by GraphEx over a period of 2 weeks, compared to the other sources of recommendations. GraphEx provides 43\% more distinct item-keyphrase associations than Graphite with the average search volume of its keyphrases nearing 30x of RE, and 2.5X of fastText. In terms of performance, GraphEx delivers an incremental lift of 8.3\% in total ads revenue and a 10.3\% in Gross Merchandise volume Bought (GMB), i.e. the total money made by selling the item. In terms of Return on Ads Spend (ROAS), given by $ROAS = \frac{GMB}{Ads\;Revenue}$, it is the most successful among the cold-start models. Among all models its ROAS is only beaten by RE which are non-cold start 100\% recall models, and GraphEx beats them in terms of item coverage (more than 3x items covered by GraphEx). We cannot disclose anymore details due to business and proprietary reasons.

\section{Conclusion}
We introduce a novel graph-based extraction method called GraphEx which is tailored for online advertising in the e-commerce sector. GraphEx efficiently solves the permutation
problem of token extraction from item's title and mapping them to a set of valid keyphrases. It is not limited by the vocabulary of the item's title and the order of tokens in them. This method produces more item-relevant keyphrases and also targets head keyphrases favored by advertisers, ultimately driving more sales. It is currently implemented at eBay, a leading e-commerce platform serving its sellers with billions of items daily. We show that traditional metrics do not provide accurate comparison amongst the models, and using a single metric for comparison will be misleading. Thus, we use a combination of metrics with AI evaluations to provide a better picture of the practical challenges of keyphrase recommendation. We evaluated its performance against the production models on eBay, demonstrating superior results for our model across the various metrics. Additionally, GraphEx offers the most profitable cold start keyphrase recommendations for advertisers with the lowest inference latency in eBay's current system and allows for daily model refreshes to serve our ever-changing query space.

\newpage
\bibliographystyle{IEEEtran}
\bibliography{refs}

\end{document}